\documentstyle[12pt,epsfig]{cernart}
\begin{document}
\begin{titlepage}


\title{\Large One and two dimensional analysis of 3$\pi$ correlations measured in Pb+Pb interactions}

\begin{Authlist}
{
I.~G.~Bearden$^{1}$, H.~B{\o}ggild$^{1}$, J.~Boissevain$^{2}$, 
P.~Christiansen$^{1}$, L. Conin$^{3}$, 
J.~Dodd$^{4}$, B.~Erazmus$^{3}$, S.~Esumi$^{5,a}$, C.W.~Fabjan$^{6}$, 
D.~Ferenc$^{7}$, D.~E.~Fields$^{2,b}$, A.~Franz$^{6,c}$, 
J.~Gaardh\o je$^{1}$, A.~G.~Hansen$^{1}$, 
O.~Hansen$^{1}$, D.~Hardtke$^{8,d}$, H.~van~Hecke$^{2}$, 
E.~B.~Holzer$^{6}$, T.J.~Humanic$^{8}$, P.~Hummel$^{6}$, B.V.~Jacak$^{9}$, 
R.~Jayanti$^{8}$, K.~Kaimi$^{5,\dagger}$, M.~Kaneta$^{5}$, T.~Kohama$^{5}$, 
M.~Kopytine$^{9,e}$, M.~Leltchouk$^{4}$, A.~Ljubi\v ci{\'c}, Jr.$^{7,c}$, 
B.~L{\"o}rstad$^{10}$, N.~Maeda$^{5,f}$, 
L. Martin$^{3}$, A.~Medvedev$^{4}$, M.~Murray$^{11}$,  
H.~Ohnishi$^{5,c}$, G.~Pai\'c$^{6,8}$, S.U.~Pandey$^{8,g}$,
F.~Piuz$^{6}$, J.~Pluta$^{3,h}$, V.~Polychronakos$^{12}$, M.~Potekhin$^{4}$,
G.~Poulard$^{6}$, D.~Reichhold$^{8}$, A.~Sakaguchi$^{5,i}$,
J.~Schmidt-S{\o}rensen$^{10}$, J.~Simon-Gillo$^{2}$, W.~Sondheim$^{2}$, 
T.~Sugitate$^{5}$, J.P.~Sullivan$^{2}$, Y.~Sumi$^{5,j}$,
W.J.~Willis$^{4}$,
K.~L.~Wolf$^{11,\dagger}$, N.~Xu$^{2,d}$, and D.~S.~Zachary$^{8}$\\[2mm]
(NA44 Collaboration)\\[4mm]
}
\end{Authlist}

\begin{abstract}
$\pi^{-}\pi^{-}\pi^{-}$ correlations from Pb+Pb collisions at 158 GeV/c per nucleon are presented as measured by the focusing spectrometer of the NA44 experiment at CERN. The three-body effect is found to be stronger for PbPb than for SPb. The two-dimensional three-particle correlation function is also measured and the longitudinal extension of the source is larger than the transverse extension.
\end{abstract}

\submitted{Final version}

\textheight = 24cm

\begin{flushleft}
{\footnotesize
$^{1}$ Niels Bohr Institute, DK-2100 Copenhagen, Denmark.\\
$^{2}$ Los Alamos National Laboratory, Los Alamos, NM 87545, USA.\\
$^{3}$ Nuclear Physics Laboratory of Nantes, 44072 Nantes, France. \\ 
$^{4}$ Columbia University, New York, NY 10027, USA.\\
$^{5}$ Hiroshima University, Higashi-Hiroshima 739, Japan.\\
$^{6}$ CERN, CH-1211 Geneva 23, Switzerland.\\
$^{7}$ Rudjer Boskovic Institute, Zagreb, Croatia.\\
$^{8}$ Department of Physics, The Ohio State University, Columbus, OH 43210, USA.\\
$^{9}$ State University of New York, Stony Brook, NY 11794, USA.\\
$^{10}$ University of Lund, S-22362 Lund, Sweden.\\
$^{11}$ Texas A\&M University, College Station, TX 77843-3366, USA.\\
$^{12}$ Brookhaven National Laboratory, Upton, NY 11973, USA.\\
$^{a}$ Now at Universit{\"a}t Heidelberg, D-69120 Heidelberg, Germany.\\
$^{b}$ Now at University of New Mexico, Albuquerque, NM 87131, USA.\\
$^{c}$ Now at Brookhaven National Laboratory, Upton, NY 11973, USA.\\
$^{d}$ Now at Lawrence Berkely National Laboratory, Berkely, CA 94720, USA. \\
$^{e}$ on an unpaid leave from P.N.~Lebedev Physical Institute, Russian Academy of Sciences, Russia.\\
$^{f}$ Now at Florida State University, Tallahassee, FL 32306 USA. \\
$^{g}$ Now at Wayne State University, Detroit, MI 48202, USA.\\
$^{h}$ Permanent adress: Warsaw University of Technology, Koszykowa 75, 00-662 Warsaw, Poland.\\
$^{i}$ Now at Osaka University, Toyonaka-shi, Osaka 560-0043, Japan.\\
$^{j}$ Now at Hiroshima International University, Kurose-cho, Hiroshima 724-0695, Japan\\
$\dagger$ Deceased.}
\end{flushleft}
\end{titlepage}

\section{INTRODUCTION}
Three-particle Bose-Einstein correlations are sensitive to source geometry and chao-ticity\cite{3pheinz,axelphase} in a way not seen in ordinary two-particle correlations. Hence three-body correlation studies provide a test of the validity of various assumptions often used in the parameterization of the two-particle correlation function. 
The additional information is carried by the genuine three-particle correlation term. In the case of source asymmetries and/or coherent particle emission, the strength of the genuine three-particle correlation term will be suppressed. 
Genuine three-particle correlations have been found at their full strength in electron-positron annihilations\cite{opal} and in high energy in $p\overline{p}$ collisions\cite{ua1}. In other reactions and at lower energies, the possible presence of a three-body correlation is unclear\cite{na22,ISR87,bock92,wolf}. 
In heavy-ion collisions a significant suppression of the genuine three-body correlation has been 
reported by this collaboration in S+Pb collisions\cite{NA4499a}. In this paper we have analysed our Pb+Pb three-pion events in the same way as in \cite{NA4499a}. In addition we report on the first experimental measurement of the two-dimensional three-particle correlation function.

\section{EXPERIMENTAL SET-UP}

The NA44 experiment is a focusing spectrometer measuring particle distributions at mid-rapidity with excellent particle identification. The spectrometer set-up has been described well in \cite{NA4498a}.
The momentum range and sign of the charge of the particles is selected by two dipole magnets. The data used for this analysis is 4 GeV/c $\pm$ 20\%. The spectrometer axis is located at 44 mrad with respect to the beam axis and covers a $p_T$ range of 0 - 400 MeV/c, with an average of $\langle p_T \rangle$ = 145 MeV/c. The rapidity range is $3.1-4.1$, with an average of 3.7. 
The trigger requires a well identified single lead ion, at least two hits on the hodoscopes and a pion signal (with no electrons) from the Cherenkov detectors. The centrality is fixed at the most central 9\% of the geometrical
cross-section by means of a threshold on a scintillator downstream of the target.

\section{THE PARAMETERIZATION OF THE CORRELATION FUNCTIONS}
In the case of a totally chaotic source, the two- and three-particle Bose-Einstein correlation functions can  be written as\cite{NA4499a}:
\begin{eqnarray}
C_2 &=& 1 + |F_{ij}|^2 \label{c2},\\
C_3 &=& 1 + |F_{12}|^2 + |F_{23}|^2 + |F_{31}|^2 + 2Re(F_{12}F_{23}F_{31} ),
\label{c3}
\end{eqnarray}
where, assuming plane wave propagation:  
\begin{equation}
F_{ij} \equiv \int e^{iQ_{ij}r} \rho(r) d^4r,~~~~~~~~~~~~ij=12,23,31
\label{fourier}
\end{equation}
where $Q_{ij}$ is the four-momentum difference of particle $i$ and $j$ and $\rho$ is the source density function. The last term in Eq. (\ref{c3}) is the so-called genuine three-particle correlation due to a pure three-body effect.
By assuming a symmetric Gaussian source density function $\rho$ of width $R$ the correlation functions are parametrized as:
\begin{eqnarray}
C_2(Q_{ij}) &=& 1+ \lambda e^{-Q_{ij}^2R^2} \label{cor-function2}, \\
C_3(Q_{12},Q_{23},Q_{31}) &=& 1 + \lambda \sum_{ij} e^{-Q_{ij}^2R^2} 
+ 2 \lambda^{3/2} e^{-\frac{1}{2}(\Sigma_{ij}Q_{ij}^2)R^2},
\label{cor-function3}
\end{eqnarray}
where $ij$ is one of the permutations $ij=12,23,31$. 
Here $\lambda$ is a phenomenological\cite{deutchmann} parameter defined by: $\lambda \equiv C(Q=0)-1$. This parameter has been introduced due to the fact that the measured two- and three-particle correlation functions do not reach the full value which, for Bose-particles, are 2 and 6 for two and three particles respectively. 

We have analysed the three-particle correlation function in terms of the sum of the momentum differences, $Q_3$, and the components of $Q_3$ transverse and along the beam, i.e. $Q_3^2=Q_t^2+Q_l^2$. We will use Eqs. (\ref{cor-function2}), (\ref{dgaussian}), and (\ref{c32d-para}) to parameterize our data:
\begin{equation}
C_3(Q_3)=1+\lambda_3e^{-Q_3^2 R_3^2},
\label{dgaussian}
\end{equation}
where $Q^2_3 = Q^2_{12} + Q^2_{23} + Q^2_{31}$, and:
\begin{equation}
C_3(Q_t,Q_l)=1+\lambda_{2d} e^{-(R_t^2 Q_t^2 + R_l^2 Q_l^2 )},
\label{c32d-para}
\end{equation}
calculated in the longitudinal center-of-mass system for the triplets characterized by $\sum p_z=0$. The kinematical variables used in Eq. (\ref{c32d-para}) are defined in Eqs. (\ref{c32d-t}-\ref{c32d-l}):
\begin{eqnarray}
Q_t &=&\sqrt{q_{T,12}^2+q_{T,23}^2+q_{T,31}^2}, ~~~~q_{T,ij}^2=(p_{x,i}-p_{x,j})^2+(p_{y,i}-p_{y,j})^2 \label{c32d-t} \\
Q_l &=& \sqrt{q_{L,12}^2+q_{L,23}^2+q_{L,31}^2}, ~~~~q_{L,ij}^2=(p_{z,i}-p_{z,j})^2 \label{c32d-l} 
\end{eqnarray}
where $p_x$, $p_y$, and $p_z$ are the three momenta components.

\section{THE GENUINE THREE-PARTICLE CORRELATION}
We introduce a weight factor $\omega$ defined by the relation:
\begin{eqnarray} 
& C_3 = 1 + |F_{12}|^2 + |F_{23}|^2 + |F_{31}|^2 +  
2|F_{12}||F_{23}||F_{31}| \times \omega.
\label{c4} 
\end{eqnarray} 
The weight factor $\omega$ is a measure of the strength of the genuine 
three-particle correlation. It can be experimentally determined by using the 
following expression, extracted from Eq. (\ref{c4}):
\begin{equation} 
\omega = \frac{\{C_3(Q_3)-1\}-\{C_2(Q_{12})-1\}-\{C_2(Q_{23})-1\}-\{C_2(Q_{31})
-1\} }  
{2\sqrt{\{C_2(Q_{12})-1\}\{C_2(Q_{23})-1\}\{C_2(Q_{31})-1\}}}.
\label{eq:phase1} 
\end{equation} 
For a totally chaotic and symmetric source $\omega=1$, but $\omega$ will 
differ from 1 for an asymmetric and/or coherent particle emitting source, see \cite{NA4499a,3pheinz,axelphase,seki99a,seki99b}.

Traditionally, the chaoticity is measured by the intercept parameter $\lambda$ of the two-particle correlation function. In practice this intercept measurement is difficult since phase space approaches 0 as $Q_{ij} \rightarrow 0$ and $C_2$ is  diluted by resonances and smeared out by the effects of momentum resolution and coulomb repulsion. The $\omega$ is, however, calculated over a relatively broad range in $Q$. Any deviation of $\omega$ from $1$ is an indication of coherence as asymmetry effects have only minor influence on $\omega$\cite{axelphase}. Note that lambda cancels in Eq. (\ref{eq:phase1}). This is seen directly for $Q_{ij}=0$ as $\omega(Q=0)=1$ for all values of lambda.

\section{DATA ANALYSIS}

We study negative pions produced when a lead beam hits a lead target. The dataset consists of 94,000 events with three pions. We find tracks by fitting straight lines to hits on two hodoscopes, two strip chambers and a pad chamber, all situated behind the magnets.
Tracks are not allowed to share the same hodoscope slats and a minimum track separation is required in the pad-chamber to certify a high purity of the three-pion sample. 
The time-of-flight start signal is derived from a beam counter with a time resolution of $\sigma \simeq$ 35 ps\cite{NA4492c}. Events containing particles other than pions are rejected by 
combining information from the time-of-flight and from the multi-particle threshold imaging Cherenkov (TIC) detector\cite{tic}. The TIC distinguishes pions from heavier particles on a track by track basis by requiring a sufficiently high signal in a fiducial zone around the tracks. 
The residual contamination from particles other than pions is typically less than 1\%. 
Identical cuts are used in constructing the two- and three-particle correlation functions. 

The bin size used in the analysis corresponds to the momentum resolution. The resolutions are calculated using a Monte-Carlo programme with full detector simulation. 

The correlation function is determined using $C(\vec{Q}) = A(\vec{Q})/B(\vec{Q})$. The ``real'' momentum distribution $A(\vec{Q})$ is constructed from tracks from the same event and 
the ``background'' distribution, $B(\vec{Q})$, is constructed from tracks mixed randomly from all events contained in $A(\vec{Q})$. Around ten ``background'' events 
are created for each ``real'' event in order to avoid statistical uncertainties from the ``background'' sample. This method cancels effects of the experimental acceptance and trigger biases and is described in detail in our previous publication\cite{na4494b}.

To compare with theoretical correlation functions, corrections are applied iteratively to produce the correlation function $C_{corr}$:
\begin{equation}
C_{corr} =C_{raw} \times K_{SPC} \times K_{acceptance} \times K_{Coulomb}.
\end{equation}
This procedure converges within four iterations. The factors are explained below:

$\bullet$ The background spectrum is distorted with respect to the true uncorrelated many-particle spectrum, owing to the effect of the many-particle correlations on the single-particle spectrum. This is iteratively corrected by the factor $K_{SPC}$ appropriately generalized to the three-particle case, in which each particle used in the background spectrum is weighted by the correlation from the event from which it is taken.

$\bullet$ The factor $K_{acceptance}$ corrects the data for the momentum resolution of the spectrometer and the many-particle acceptance and is calculated using a Monte Carlo programme with a full simulation of the tracking detectors and multiple scattering.

$\bullet$ The two-particle correlation function has been corrected for the Coulomb interaction, $K_{Coulomb}$, between the particles using the Coulomb wave function integration method\cite{Pra86c,biyajima}. The three-particle correlation function has been corrected for the Coulomb interaction using a similar technique\cite{threecoulomb, NA4499a} by using a three-body Coulomb wave-function\cite{merkuriev,brauner}.

Coulomb interactions with the residual nuclear system are expected to be small and no corrections are applied in this analysis. Final-state strong interactions are also expected to be small and due to large uncertainties in proposed procedures, no corrections are applied for them\cite{Boa90a}. 

The systematic errors are evaluated by varying the analysis parameters and calculating the difference in the correlation functions produced. These variations include changing the momentum resolution assumed in the Monte-Carlo correction by $\pm10$\%, changing the time-of-flight cuts, 
increasing the minimum separation in the pad chamber, requiring more hits in the strip chambers, and increasing the minimum slat separation in all the hodoscopes. We re-calculate $\omega$ for each new set of correlation functions. 
The systematic errors are estimated by summing up the differences to the mean-value for each altered setting. The statistical error on the $\omega$ distribution is: $\sigma_{stat}(\overline{\omega})=\sigma/\sqrt{N}$, where $\sigma$ is the variance and $N$ is the number of entries. The statistical error is much smaller than the systematic error. In the statistical error calculation we have not included error propagation from the correlation functions to $\omega$. This is very complicated so instead we have calculated $\omega$ using $C_3 \pm \sigma_{stat}$ and $C_2 \pm \sigma_{stat}$ and treated the deviation as an additional systematic error. The two systematic errors are added in quadrature.

There is still some room for further systematic errors in the Coulomb correction, due to the fact that the unknown exact three-body Coulomb wave-function is reproduced only asymptotically. However, in the kinematic region of the NA44 experiment, these non-asymptotic correction terms in the three-body Coulomb wave-function are known to decrease strongly with increasing energy of the triplets and the average energy of the triplet is large compared to the typical scale of the three-body Coulomb potential in the NA44 three-pion data sample, see \cite{threecoulomb}.

\section{RESULTS AND DISCUSSION}

\begin{figure}
\begin{minipage}[h]{75mm} 
\epsfig{file=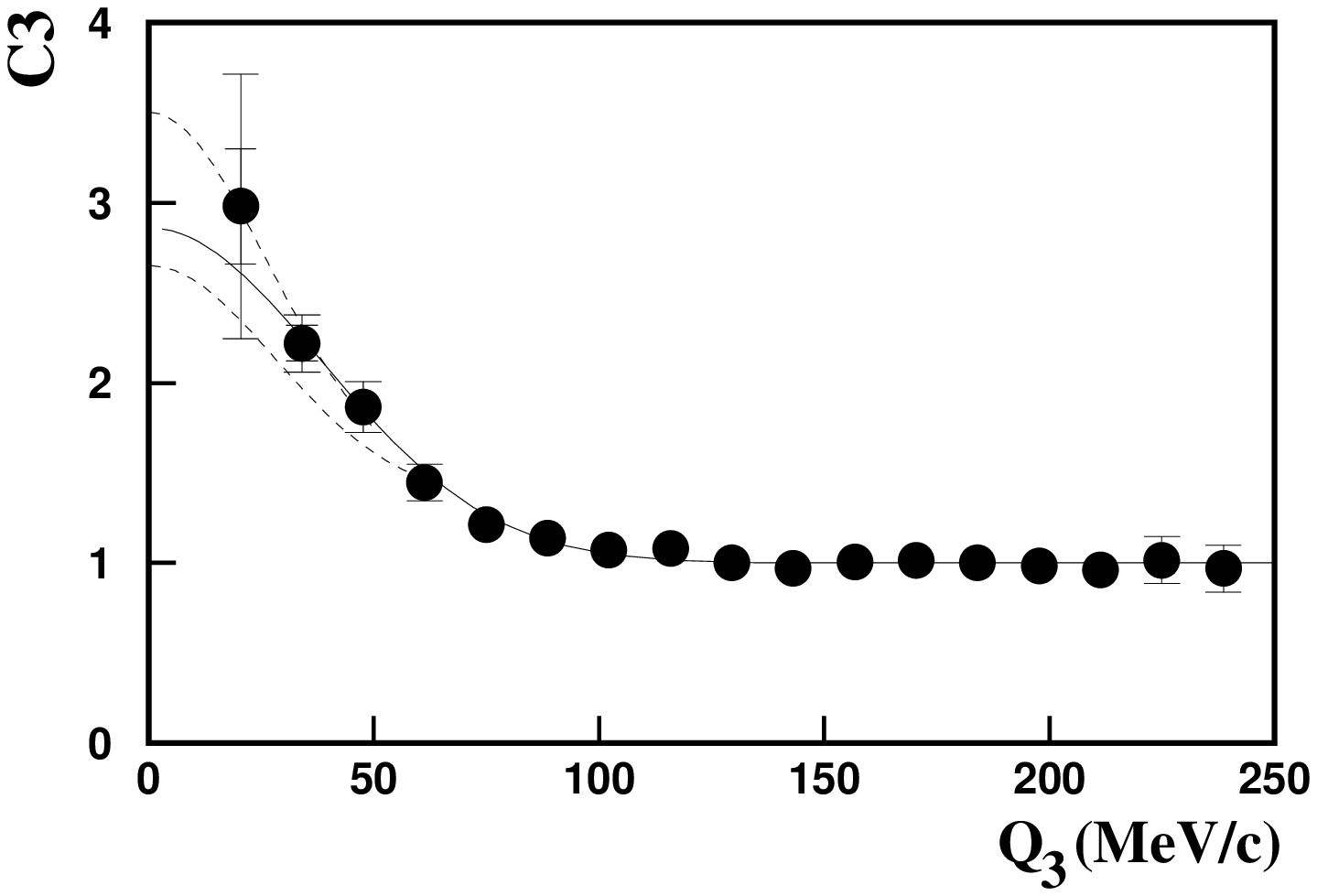,width=7cm}
\caption{The three-particle correlation function fitted (solid line) using Eq. (\ref{dgaussian}). The double error bars are statistical (inner) and systematic (outer) respectively. The dotted lines indicate what path the datapoints should follow with (upper) and without (lower) the genuine three-particle correlation, see the text.}
\label{fg:c3}
\end{minipage} 
\hspace{\fill} 
%
\begin{minipage}[h]{75mm} 
\epsfig{file=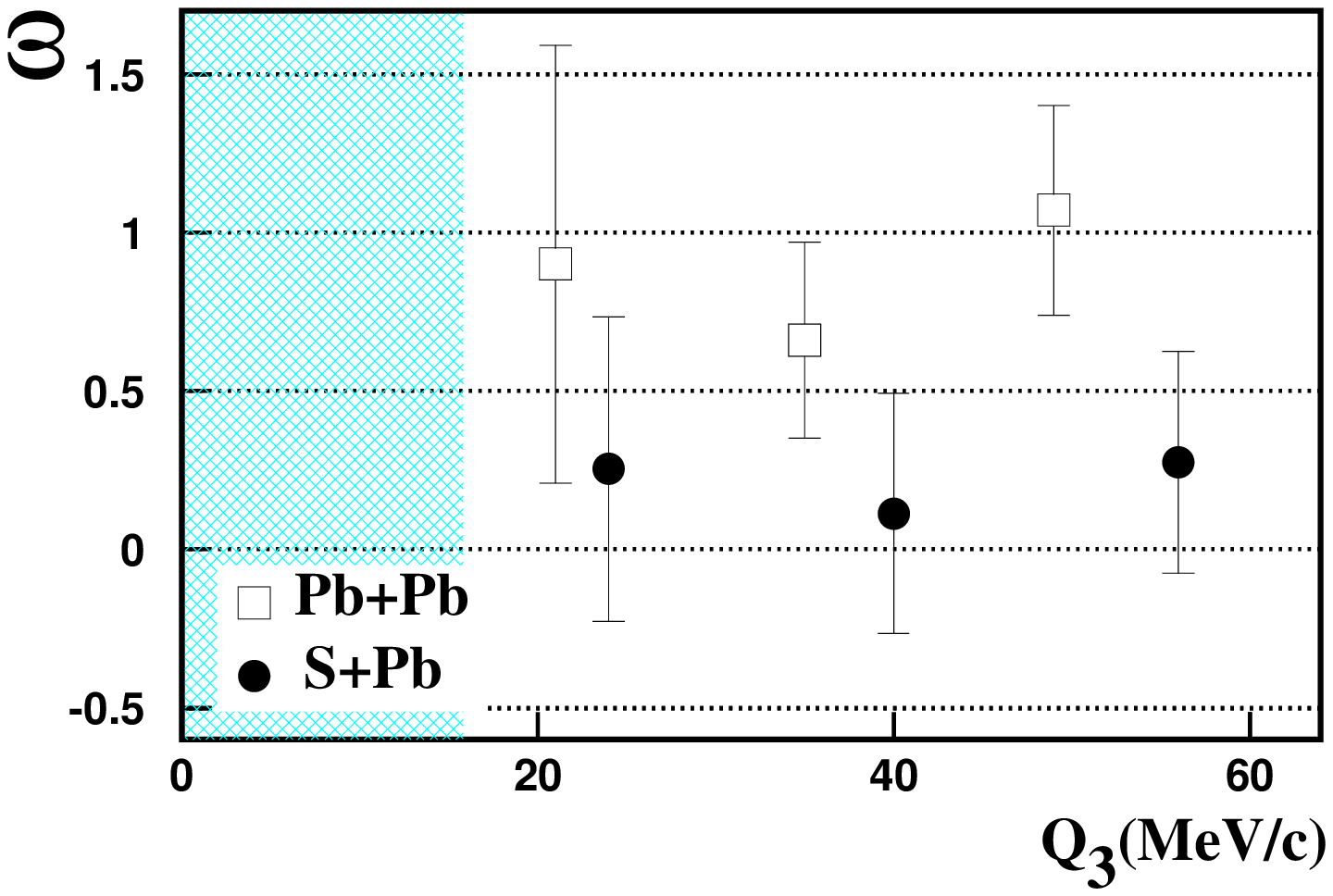,width=7cm}
\caption{The weight factor $\omega$ as a function of $Q_3$ for Pb+Pb (open box) and S+Pb (closed circles). The dashed area indicates the $Q_3$-range without data. The errors shown are systematic only since the statistical ones are within the data points. Bear in mind that the sytematic errors are correlated within each data set as well as for the two data sets (see text).}
\label{fg:pic5}
\end{minipage} 
\end{figure}

The three-pion correlation function is shown in Fig. \ref{fg:c3} and the result of a fit to Eq. (\ref{dgaussian}) is summarized in Table~\ref{tb:1}. An estimate of the strength of the three-pion correlation function is shown in Fig. \ref{fg:c3} as dotted lines. This is done using Eq. (\ref{cor-function3}) with $\lambda$ and $R$ extracted from a fit to the two-pion correlation function. The lower dotted 
line is without contribution from the genuine three-body correlation, i.e. the datapoints would follow this path in case of $\omega$=0. The upper dotted line is for $\omega$ =1, and in fact seems to follow the datapoints. 
\begin{table}
\centering
\begin{tabular}{|c|c|c|} \hline
$\lambda_3$ & $R_3$ (fm) & $\chi^2$/N$_{dof}$ \\ \hline
1.92 $\pm$ 0.16 $\pm$ 0.33 &  3.78 $\pm$ 0.16 $\pm$ 0.37 &  17.1/15 \\ \hline
\end{tabular}
\caption{Results from a fit to the three-pion correlation function using Eq. (\ref{dgaussian}). The errors are statistical and systematic respectively. }
\label{tb:1}
\end{table}

The two-pion correlation function is constructed by using the three combinations of a pair from the triplet datasample. The correlation function is fitted using Eq. (\ref{cor-function2}) and the results are summarized in Table \ref{tb:2} where we also list our results from \cite{NA4498a}. 
The extracted fit-parameters are consistent between the two- and three-particle datasets despite the different centrality. The similarity indicates that the effect on the correlation function due to the presence of a third pion is small for our data.
By comparing results in Table \ref{tb:2} and \ref{tb:1} one sees directly that $\lambda_3 > 3\lambda$ as expected when $\omega > 0$. 

\begin{table}
\centering
\begin{tabular}{|l|c|c|c|c|} \hline
System & $\lambda$ & $R$ (fm) & $\chi^2$/N$_{dof}$ & Centrality \\ \hline
 $3\pi \rightarrow 2\pi$ & 0.57 $\pm$ 0.04 & 7.49 $\pm$ 0.34 & 18/20 & 9\% \\ \hline 
 $2\pi$\cite{NA4499a} & 0.52 $\pm$ 0.04  & 7.56 $\pm$ 0.38 & 30/36  & 18\% \\ \hline 
\end{tabular}
\caption{Results from a fit to the two-pion correlation function made from the three- and a two-particle dataset. Both datasets are corrected by the Coulomb-Wave integration but they have different centrality. The errors are statistical only.}
\label{tb:2}
\end{table}
The direct method of extracting the strength of the genuine three-pion correlation is to calculate the $\omega$ factor using Eq. (\ref{eq:phase1}). 
When we have determined $C_2$ and $C_3$, as described above, we can use the actual data points, $C_2(Q_{12})$, $C_2(Q_{23})$, $C_2(Q_{31})$, and $C_3(Q_3)$, so as not to be biased by some parameterization. The data points are obtained by using $Q_{12}$, $Q_{23}$, $Q_{31}$, and $Q_3$ for each event and we have checked that the result do not change when altering the $Q$'s by $\pm$ 5\%. In order to avoid poles in the denominator of Eq. (\ref{eq:phase1}) events are accepted if $Q_{ij} \le 60$ MeV/c. As a result we obtain a distribution of $\omega$ for each $Q_3$ bin, see Fig. \ref{fg:pic5}, where we also show our S+Pb result\cite{NA4499a}. Events are taken in the region of the genuine three-body correlation, i.e. in the range $14 < Q_3 < 54$ MeV/c as we have no data below $14$ MeV/c. In this range we find the weighted-mean $\overline{\omega}$ = 0.85 $\pm$ 0.02 $\pm$ 0.21 where the errors are statistical and systematic, respectively. The value $\overline{\omega}$ from our S+Pb result\cite{NA4499a} was $\overline{\omega}$ = 0.20 $\pm$ 0.02 $\pm$ 0.19.

The three-pion correlation in PbPb collisions has also been published by WA98\cite{wa98} without using the more correct Coulomb correction method\cite{threecoulomb} employed by NA44. Within the errors, however, the $\overline{\omega}$ are compatible between the two experiments.

\begin{figure}[hbtp]
\centering
\epsfig{file=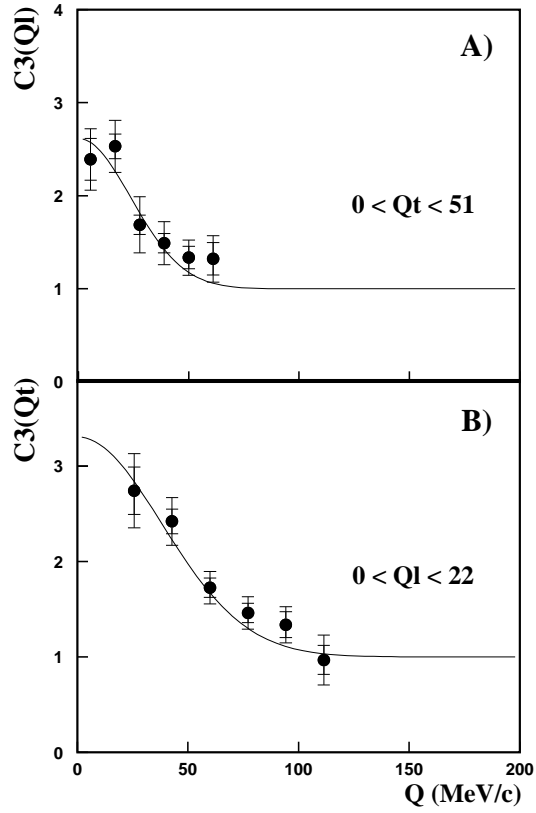,width=8cm}
\caption{The two-dimensional three-pion correlation function is projected onto the A) $Q_l$ and B) $Q_t$ axis for a range in $Q_t$ and $Q_l$ as indicated in the figure. The solid line is drawn using Eq. (\ref{c32d-para}) using an average $\langle Q \rangle$ in the orthorgonal direction of the momentum difference. The correlation function is calculated in the longitudinal center-of-mass system for the triplets. 
The errors are statistical (inner) and systematic (outer) respectively.}
\label{fg:c32d}
\end{figure}

The two-dimensional three-pion correlation function has also been analyzed in the longitudinal-center-of-mass system. Only bins which contain more than 40 entries are used in the fit. Projections are made following \cite{Cha91a} and are shown in Fig. \ref{fg:c32d}. The results of a fit to the two-dimensional correlation function using Eq. (\ref{c32d-para}) is summarized in Table~\ref{tb:1b}. The systematic error of $R_t$ and $R_l$ are correlated, i.e. the radii both become small or both large when varying the analysis parameters. When comparing the difference of the radii to the largest statistical error we see a 4-$\sigma$ effect in the difference of the radii.

There exist only few predictions of the two-dimensional three-body correlation function. One is based on the Lund string-model\cite{markus} where a larger longitudinal than transverse radius parameter is in fact expected.
\begin{table}
\centering
\begin{tabular}{|c|c|c|c|} \hline
$\lambda_{2d}$ & $R_t$ (fm) & $R_l$ (fm) & $\chi^{2}$/N$_{dof}$ \\ \hline
2.67 $\pm$ 0.49 $\pm$ 0.19 &  3.67 $\pm$ 0.26 $\pm$ 0.19 & 5.89 $\pm$ 0.53 $\pm$ 0.68 &   189/113 \\ \hline
\end{tabular}
\caption{Results of a fit to the two-dimensional three-pion correlation function using Eq. (\ref{c32d-para}). The errors are statistical and systematic respectively. }
\label{tb:1b}
\end{table}

\section{CONCLUSIONS}

We have found that the measure of the strength of the genuine three-particle correlation, expressed as the mean weight factor $\omega$ in PbPb interactions, is compatible with 1, $\omega$= 0.85 $\pm$ 0.02 $\pm$ 0.21. This is different from SPb interactions where we earlier have found, using the same analysis method, $\omega$=0.20 $\pm$ 0.02 $\pm$ 0.19. We consider this difference to be significant as the systematic errors between the two dataset are correlated. The small $\omega$-value in SPb interactions indicates a nonchaotic mechanism for particle production\cite{3pheinz,axelphase,seki99d,seki99c}, different from  PbPb interactions which are compatible with a fully chaotic mechanism.\\

We have performed a first measurement of the two-dimensional three-pion correlation function. The data show that the longitudinal radius parameter is larger than the transversal radius parameter in the longitudinal center-of-mass system.

\section{ACKNOWLEDGEMENTS}

The NA44 Collaboration wish to thank the staff of the CERN PS-SPS
accelerator complex for their excellent work.
We thank the technical staff at CERN and the collabo-rating institutes for
their valuable contribution.
We are also grateful for the support given by
the Science Research Council of Denmark; 
the Austrian Fond f{\"u}r F{\"o}rderung der Wissenschaftlichen Forschung through grant P09586;                       
the Japanese Society for the Promotion of Science, and
the Ministry of Education, Science and Culture, Japan;
the Science Research Council of Sweden;
the US Department of Energy; and the National 
Science Foundation.
 
\bibliography{./NA44abrv,./three}
\bibliographystyle{./NA44usrt}         
 

\end{document}